\documentclass[submission,copyright,creativecommons]{eptcs}
\providecommand{\event}{MARS 2015} 
\usepackage{breakurl}             

\usepackage{isabelle,isabellesym} 
\isabellestyle{it} 

\usepackage[normalem]{ulem} 
\usepackage{stmaryrd} 
\usepackage{bbding} 
\usepackage{amssymb} 
\usepackage{units} 
\usepackage{color} 
\usepackage{endnotes} 
\usepackage{graphicx} 
\usepackage{xspace} 
\usepackage{appendix} 

\renewcommand\sectionautorefname{{\S}\kern-.3em} 
\renewcommand\subsectionautorefname{{\S}\kern-.3em} 
\renewcommand\figureautorefname{{Fig.\kern-.1em}} 
\newcommand\OG {\textit{OG}}
\newcommand\OS {\textit{OS}}

\newcommand\VC {\textit{VC}}

\newcommand\While {\mbox{${\mathit WHILE}$}}
\newcommand\Await {\mbox{${\mathit AWAIT}$}}
\newcommand\Then {\mbox{${\mathit THEN}$}}
\newcommand\End {\mbox{${\mathit END}$}}
\newcommand\Do {\mbox{${\mathit DO}$}}
\newcommand\Gets {\kern.2em{:=}\kern.5em}
\newcommand\True {\mbox{$\mathit{True}$}}
\newcommand\False {\mbox{$\mathit{False}$}}
\newcommand\DSemi {;} 

\DeclareMathAlphabet{\mathitbf}{OT1}{cmr}{bx}{it}

\newcommand{\AT}{\mathit{AT}}
\newcommand{\ATstack}{\mathit{ATstack}}
\newcommand{\EIT}{\mathit{EIT}}
\newcommand{\EITstack}{\mathit{EITstack}}

\newcommand{\U}{\mathrm{U}}
\newcommand{\UU}{\mathcal{U}}
\newcommand{\I}{\mathrm{I}}
\newcommand{\II}{\mathcal{I}}
\newcommand{\T}{\mathrm{T}}
\newcommand{\SVCs}{\mathrm{SVC_s}}
\newcommand{\SSS}{\mathcal{S}}  
\newcommand{\SVCa}{\mathrm{SVC_a}}
\newcommand{\AAA}{\mathcal{A}}  

\newcommand{\R}{R}  
\newcommand{\E}{E}  

\newcommand{\control}{ {\mathitbf{control}}}
\newcommand{\ITake}{ {\mathitbf{ITake}}}
\newcommand{\IRet}{ {\mathitbf{IRet}}}
\newcommand{\ITakeSVCa}{ {\mathitbf{ITakeSVC_a}}}

\newcommand{\eChronos}{\textit{eChronos}\xspace}

\newcommand{\theeChronosOS}{the \textit{eChronos} OS\xspace}
\newcommand{\TheeChronosOS}{The \textit{eChronos} OS\xspace}

\newcommand{\titl}{
Controlled Owicki-Gries Concurrency:
Reasoning about the Preemptible 
\eChronos Embedded Operating System
}
\newcommand{\runtitl}{
Controlled \OG\ Concurrency:
Reasoning about the Preemptible 
\eChronos Embedded OS
}

\title{\titl
}
\author{
June Andronick
\institute{NICTA and UNSW}
\email{june.andronick@nicta.com.au}
\and
Corey Lewis
\institute{NICTA}
\email{corey.lewis@nicta.com.au}
\and
Carroll Morgan
\institute{NICTA and UNSW}
\email{carroll.morgan@unsw.edu.au}
}
\def\titlerunning{\runtitl}
\def\authorrunning{June Andronick, Corey Lewis, Carroll Morgan}
\begin{document}
\maketitle

\begin{abstract}
We introduce a controlled concurrency framework, derived from the
\emph{Owicki-Gries} method, for describing a hardware interface in
detail sufficient to support the modelling and verification of small,
embedded operating systems (\OS's) whose run-time responsiveness is
paramount.  Such real-time systems run with interrupts mostly
enabled, including during scheduling. That differs from many other
successfully modelled and verified \OS's that typically reduce the
complexity of concurrency by running on uniprocessor platforms and by
switching interrupts off as much as possible.

Our framework builds on the traditional Owicki-Gries method, for its
fine-grained concurrency is needed for high-performance system code. We
adapt it to support explicit concurrency control, by providing a
simple, faithful representation of the hardware interface that allows
software to control the degree of interleaving between user code, \OS\
code, interrupt handlers and a scheduler that controls context
switching.  
We then apply this framework to model the interleaving behavior of
\theeChronosOS, a preemptible real-time \OS\ for embedded
micro-controllers.
We discuss the accuracy and usability of our approach when
instantiated to model \theeChronosOS.
Both our framework and the \eChronos model are formalised in the
Isabelle/HOL theorem prover, taking advantage of the high level of
automation in modern reasoning tools. 
\end{abstract}

\section{Introduction}

\emph{Formal verification} is an inescapable requirement in cases
where software/hardware failure would be catastrophic.
Existing modelled and verified operating systems
(e.g.~\cite{Klein_AEMSKH_14, Yang_Hawblitzel_10, Huang_11,
  Feng_SGD_09}) typically run on uniprocessor platforms. They are also \emph{not 
  preemptible}, i.e. they run with interrupts mostly disabled, at
least during scheduling; thus their execution is mostly {\em
  sequential}.

Here, in contrast, we target {\em preemptible} (still uniprocessor) real-time
  \OS\ code.  Our motivating example is \theeChronosOS~\cite{eChronos},
  an open-source real-time \OS\ that provides a library of \OS\ services to
  applications, including synchronisation primitives (signals,
  semaphores, mutexes), context switching, and scheduling.  Our
  approach, however, applies to any system where the \OS\ code is
  preemptible, including scheduler code, and runs on uniprocessor
  hardware that supports nested interrupts.
While being preemptible, the \OS\ code is not {\em re-entrant}, which
means that its execution can be interrupted at any moment by an
interrupt handler servicing a hardware-device interrupt (unless that
interrupt is masked off), but its execution is resumed after the
interrupt has been handled.
In order to allow faster response time, the \OS\ is also {\em
  preemptive}, meaning that it can unilaterally take control from
application tasks.

\TheeChronosOS is used in tightly constrained devices such
as medical implants, running on
embedded
micro-controllers with no memory-protection support.
It is small and comes in many variants.
The variant we are targeting (which we will from now on simply refer
to as \theeChronosOS) runs on ARM uniprocessor hardware.
\footnote{We specifically target an ARM Cortex-M4 platform, which, for
  the purposes of this paper, we will simply refer to as ARM.}
It makes use of ARM's {\em supervisor call} ({\em SVC})
mechanisms to run its scheduler, where an SVC is a program-initiated
interrupt, triggered by the execution of the SVC instruction, that
results in the execution switching to an \OS-provided SVC handler.

Earlier work has produced an initial formal specification of the
  \eChronos API, but assumed that the execution of each API function 
was sequential, i.e.\ assumed that execution of interrupt handlers could
  not affect the API's functionality.
Furthermore, it could not model the effect of context switching, 
which made proving refinement between this model and the (existing) implementation impossible.

That is what motivated the work presented here, where we 
  focus on the interleaving behavior induced by unpredictable device
  interrupts and (predictable) context switching, but still provide a
  detailed, faithful model of the precise interleaved execution of
  user tasks, SVC handlers, and interrupt handlers, including nested
  ones. For wider usability we dissociate the general controlled-concurrency framework, and formal model of the API of the hardware
  mechanisms, from its specific instantiation to the model of
  \theeChronosOS. We plan to then prove that this restricted but
  faithful model of \theeChronosOS is refined by its implementation,
  and enrich the model with a complete specification of the API.

We follow the foundational Owicki-Gries (\OG) concurrency
method~\cite{Owicki_Gries_76}, where Hoare-style assertional reasoning
is adapted to reason about a number of individually sequential
processes that are executed collectively in parallel: the execution of
the overall system is a non-deterministic 
interleaving of atomic statements each executed in the order
determined by the process within which it occurs. 
Our choice of \OG\ over more recent, derived concurrency styles, comes
from the low-level of abstraction, needed for high-performance
shared-variable system code.
We model the \OS\ system as the parallel composition of various user
tasks (consisting of application code and calls to \OS\ code), the
interrupt handlers and the SVC handlers. 
On a uniprocessor platform, this allows much more interleaving that
can happen in reality, where interleaving is controlled via hardware
mechanisms such as context switching, enabling and disabling
interrupts, etc. We adapt \OG\ by adding an explicit
control of interleaving, and we provide a formal hardware interface
for operations manipulating allowed interleaving. 
We have formalised this framework in the
Isabelle/HOL~\cite{Nipkow_PW:Isabelle} theorem prover, building on an
existing formalisation of \OG\ \cite{Prensa:PhD}.

In summary, we present the following contributions in this paper: (1)
an adaptation of the \OG-based concurrency model that controls interleaving; (2) a
concise formal model of the API of the hardware mechanisms that control the
interleaving induced by interrupts, SVC's and preemption; 
and (3) a model of the scheduling behavior of \theeChronosOS. All
of our work is formalised in Isabelle/HOL.

\section{Explicit concurrency control in Owicki-Gries reasoning}
\label{sec:OG}

The formalism we choose to
represent interleaved execution 
in the small preemptible \OS\ we aim to model,
is based on the Owicki-Gries method,
which we adapt to support explicit concurrency control.

The \OG\ method extends Hoare logic for sequential
programs~\cite{Hoare_69} to concurrent programs that share data. An
\OG\ system comprises a number of {\em tasks} built from atomic
statements. The concurrency between the tasks, i.e.\ an interleaving
of atomic executions, is essentially uncontrolled except for the {\em
  await statement} with which a task can ensure its execution is
suspended until a condition (of its choice) holds.
Await statements are of the form ${\Await}~C~{\Then}~P~{\End}$ for
some Boolean expression $C$ in the system variables and some program
fragment $P$: execution of $P$ cannot occur unless $C$ is (atomically)
evaluated to true, in which case $P$ is executed (also atomically)
immediately afterward.

An \OG\ proof generates verification conditions, \VC's, of two kinds:
conventional post-then-pre conditions, and \emph{interference-freedom}
\VC's. The latter express that one task does not falsify,
i.e.\ ``interfere with'', some conventional assertion in another task;
it is essentially a non-compositional technique, however. Worse, those
\VC's are quadratically numerous in the size of the program, which
historically has limited \OG's applicability to small systems.

Variants and extensions of \OG\ include
\emph{rely-guarantee}~\cite{Jones_83} which addresses both
compositionality and the number of \VC's. It also encourages a higher
level of abstraction, which can sometimes impose execution-time
inefficiency that might be intolerable in a high-performance
application like the real-time preemptible \OS\ we target
here. (Compare while-loops, and invariant reasoning, with super-high
performance low-level code that uses goto's in a less-structured way:
sometimes --happily, not often-- the latter is a necesary evil.)  The
same abstraction/performance tradeoff contra-indicates the use of more
structured run-time mechanisms like monitors and critical regions
\cite{Hansen_72,Hoare_74}.

Targeting high-performance code is the reason why we chose
the lower-level \OG\ style. Since we aim to
{\em verify} the \OS\ systems we have modelled, we will eventually have
to deal with the explosion of number of generated \VC's.
We believe (and have initial evidence, discussed
in~\autoref{sec:eChronos model}) that for our application, and with
our extension to control interleaving, mechanical verification is
likely to overcome those difficulties.

Our extension follows from the observation that a uniprocessor \OS\ is
not truly concurrent: via interrupts and saved contexts it interleaves
its tasks' executions in a way that simulates concurrency. Since our
modelling \emph{includes} that concurrency management, i.e. it
includes the system's scheduler code, we must include the concurrency
control in our program text.
To allow an \OG\ program to {\em control} its own interleaved
concurrency, we associate a unique value with each task and we place
each \OG-atomic command in a task within an $\Await$ condition requiring a
global variable $\AT$ (``active task'') be equal to the value
associated with that task. Suppose for example we had a Task 2 whose
atomic statements were $\mathit{First};\mathit{Second};\mathit{Third}$. It
would become:

 {\small
 $ \Await~\AT{=}2~\Then~\mathit{First}~\End; \quad
   \Await~\AT{=}2~\Then~\mathit{Second}~\End; \quad
   \Await~\AT{=}2~\Then~\mathit{Third}~\End;
 $\\[.1cm]
 } 
\noindent
Now if Task 2 were to give up control explicitly to, say some Task 1 similarly treated with $\Await~\AT{=}1$ decorations,
it would simply include the (atomic) command
{\small
\(
 \Await~\AT{=}2~\Then~\AT\Gets 1~\End;
\)
}
at the appropriate point. The basic \OG\ mechanism then ensures that Task 1 continues execution from the point it last had control.
In a more sophisticated system, Task 2 might transfer control instead to a ``scheduler task'', say Task 0, which would be a loop of the form (after pre-processing)\,%

{\small
\hspace{1cm}$
 \begin{array}{l}
  \While~\True~\Do \\
  \quad \Await~\AT{=}0~\Then~t\Gets \textit{``next runnable task''}~\End; \\
  \quad \Await~\AT{=}0~\Then~\AT\Gets t~\End \\
  \End
 \end{array}
$
}

\noindent
Our extended OG framework defines the pseudo-variable $\AT$
and uses it in the style above.  The variable is {\em pseudo} 
in the sense that it does not represent a program
variable, but rather some internal state managed by the hardware.
We provide the function $\control$ that performs the automatic
  pre-processing step within Isabelle of inserting all the
  $\Await~\AT{=}\dots$ code.
It takes as input a task identifier
  $\T$ (2 in the example above) and the task's program text, and it adds
  \Await-statements with the guard $\AT{=}\T$ to each atomic command
  of the program.  The program above would become
  $\control\ 2\ \mathcal{P}$, where $\mathcal{P}$ is Task~2's
  program (here the three instructions shown).
Our controlled interleaving will ensure that the majority of
non-interference VC's will be trivial,
\footnote{A non-interference \VC\ asserts that some assertion in task
  $X$ is preserved by some statement from task $Y$. Since most
  assertions will be guarded by $\AT{=}X$, and most statements will by
  $\AT{=}Y$, many of these \VC's will have antecedent $\False$.}  so
that mechanised verification will allow automatic~discharge.

\section{Formalisation of the Hardware Interface}
\label{sec:api}

\begin{figure}[t]
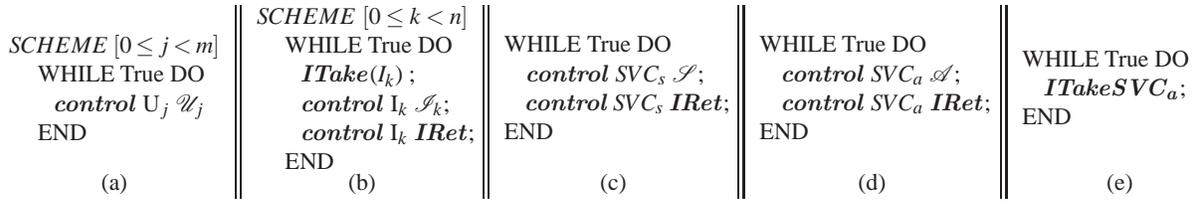

  \centering
{\footnotesize
\begin{tabular}{p{2.8cm}||p{2.8cm}||p{2.9cm}||p{3cm}||p{2.6cm}}
  \begin{minipage}{2.8cm}
                  $SCHEME\ [0 \le j < m]$\\
  \hspace*{.3cm}    WHILE True DO \\
  \hspace*{.5cm}     $\control$ $\U_j$ $\UU_j$\\
  \hspace*{.3cm}    END
  \end{minipage}
&
  \begin{minipage}{2.9cm}
                   $SCHEME\ [0 \le k < n]$\\
  \hspace*{.3cm}   WHILE True DO \\
  \hspace*{.5cm}    $\ITake$($I_k$) ;\\
  \hspace*{.5cm}    $\control$ $\I_k$ $\II_k$;\\
  \hspace*{.5cm}    $\control$ $\I_k$ $\IRet$;\\
  \hspace*{.3cm}   END
  \end{minipage}
&
  \begin{minipage}{3cm}
                  WHILE True DO \\
  \hspace*{.2cm}  $\control$ $SVC_s$ $\SSS$;\\
  \hspace*{.2cm}  $\control$ $SVC_s$ $\IRet$;\\
                  END
  \end{minipage}
&
  \begin{minipage}{3.1cm}
                  WHILE True DO \\
  \hspace*{.2cm}  $\control$ $SVC_a$ $\AAA$;\\
  \hspace*{.2cm}  $\control$ $SVC_a$ $\IRet$;\\
                  END
  \end{minipage}
&
  \begin{minipage}{2.6cm}
                  WHILE True DO \\
  \hspace*{.2cm}   $\ITakeSVCa$;\\
                  END
  \end{minipage}\\[.2cm]
\centering{(a)}&
\centering{(b)}&
\centering{(c)}&
\centering{(d)}&
\centering{(e)}
\end{tabular}\\
}
\caption{Model of OS-system with $m$ user tasks and $n$ interrupt handlers.}
\label{fig:wholesys}
\end{figure}

Our aim is to model uniprocessor, preemptible and preemptive software
systems comprising user tasks, 
interrupt handlers, and ARM-provided SVC mechanisms. The
model of~\autoref{sec:OG} represents
the interleaving between tasks. We now formalise the hardware interface that allows the \OS\ to control
interleaving. The formalisation of the whole system is then presented
in~\autoref{fig:wholesys} and explained below.

\subsection{Controlling interrupts}
\label{ssec:interrupt api}

User tasks run application code that may call
\OS\ services; and with \emph{user code} we refer to both.%
\footnote{A real-time \OS\ typically runs as an \OS\ library with no mode-switch, 
modelled here by \OS\ code being inlined in the application code.}
Assuming we have user tasks $\U_1$,~...,~$\U_m$,
the code of task $\U_j$ will be noted $\UU_j$.
The user-task part of the system is represented by (a)
in~\autoref{fig:wholesys}, where the notation $SCHEME\ [0 \le j < m]\
c_j$, borrowed from~\cite{Prensa:PhD}, is the {\em parametric}
representation of $m$ parallel processes, i.e. $ c_0\ ||\ c_2\ ||\ ...\ ||\ c_{m-1}$.

When an unmasked interrupt occurs, the running code is stopped,
context saved, and interrupt-handler code starts instead.
At the end of the execution of the interrupt handler, the
hardware performs a return-from-interrupt instruction, restoring the
saved context and switching back to the appropriate task.
The code $\II_k$ for the
interrupt task $\I_k$ might include
potential \OS-wrapper code. This interrupt part of the system is
modelled by (b) in~\autoref{fig:wholesys}.
Non-deterministic occurrence of interrupts is analogous to \OG's
spontaneous (uncontrolled) task-switching, so the hardware mechanism
that traps to the interrupt code ($\ITake$) should \emph{not} be guarded (and this is where interleaving happens). It
should model the saving of context and updating of the $\AT$ variable
to be $\I_k$. In terms of context that needs to be saved, we only
model here what is relevant to the task interleaving. The identity of
the task being interrupted (i.e. the value of $\AT$) needs to be saved
(before being updated to $\I_k$), to be able to return to
it. On platforms like ARM, nested interrupts are allowed
(i.e.\ interrupt handlers can be interrupted), so we save the whole stack of
interrupted tasks. For this, we use a second
pseudo-variable $\ATstack$. 
In reality, interrupts can be {\em masked} using specific hardware
functions, and may also obey some platform-dependent {\em interrupt
  policy}. Masked interrupts remain pending until they are
unmasked. We use a third pseudo-variable $\EIT$ to represent the set
of ``enabled interrupt tasks'', i.e. the hardware mask
bits. This set can be manipulated using the following two functions,
representing the hardware API to manipulate the interrupt mask.

\vspace{.1cm}

\noindent
\hspace*{.4cm} $Int\_Disable(X) \ \ \equiv \ \ \EIT{\Gets}\EIT-{X} $ 
\hspace{.3cm}(1)
\hspace{.8cm} and
\hspace{.8cm} $Int\_Enable(X) \ \ \equiv \ \ \EIT{\Gets}\EIT\cup{X}$
\hspace{.3cm} (2)

\noindent
These functions can be called inside \OS\ code or interrupt code.
Note that when an
interrupt is re-enabled, the hardware checks whether that interrupt occurred
while masked and is still pending, in which case it traps to the
corresponding interrupt handler. The interesting point is that our
modelling of interrupts, i.e. allowing interrupt code to
non-deterministically run at any time, already represents this case,
so adding the interrupt identifier to $\EIT$ is enough.

The interrupt policy defines the allowed nesting of interrupts. For
instance, on ARM, interrupts can only be interrupted by a higher
priority interrupt. In our model, we leave this platform-dependent
policy generic, using a predicate $interrupt\_policy~X~Y$ which is true
only if $Y$ is allowed to interrupt $X$.

The full sequence of what happens when an interrupt occurs is that the hardware
checks whether (i) the interrupt is unmasked (i.e. is in $\EIT$ in our
model), (ii) the interrupt is not the same interrupt that is being
processed (i.e. is not $\AT$ in our model), because if this happens
the interrupt will remain pending as with any masked interrupt,
(iii) the interrupt is allowed to interrupt the
currently running task, according to the interrupt policy. 
If all these conditions are met, then the context is
saved (now including the variable $\EIT$, also saved on a stack
$\EITstack$ for similar reasons as for $\AT$)
and control is switched to the interrupt handler. So in total the $\ITake$
function is defined as follows (where the operator $+$ pushes an element onto the
stack):

\vspace{.1cm}

\noindent
\hspace*{.4cm} $\ITake(X) \ \ \equiv \ \ $ 
                   $AWAIT\ X \in \EIT-\AT-\ATstack\ \wedge\ (interrupt\_policy\ \AT\ X)\ THEN$\\
\hspace*{3.4cm}        $\ATstack\Gets\AT+\ATstack; $\\
\hspace*{3.4cm}        $\EITstack\Gets\EIT+\EITstack; $\\
\hspace*{3.4cm}        $\AT{\Gets}X; $\\
\hspace*{3cm}     $END$

\noindent
Note that all statements inside an \Await\ are executed
atomically, where the atomicity here is ensured by the hardware
(i.e.\ these multiple instructions represent a single atomic hardware-defined mechanism).

By virtue of our extended \OG\ with $\textit{AWAIT}$-guarded atomic
statements, this enforces that the handler runs until further update
of the $\AT$ variable, which only happens at the return from interrupt
(or if another, unmasked, higher-priority interrupt occurs).

When returning from an interrupt $I$, the hardware will check whether there
are any pending interrupts that would have happened during the
handling of $I$ but could not run because they were masked or because
of the interrupt policy. Similarly to re-enabling interrupts, this case
is already modelled by allowing non-deterministic 
interrupts at any time. So the return-from-interrupt function $\IRet$
only needs to model the context restore (updating $\AT$ and $\EIT$
with the top of $\ATstack$ and $\EITstack$ respectively, and popping
$\ATstack$ and $\EITstack$). As opposed to $\ITake$, $\IRet$ needs to be guarded 
(using our $\control$ mechanism) as is
the rest of interrupt code, because it should only run at the end of the interrupt code.

\vspace{.1cm}

\noindent
\hspace*{.4cm} 
  $\IRet(X) \ \ \equiv \ \ $          
        $(\AT+\ATstack){\Gets}\ATstack;$ $\ \ $
        $(\EIT+\ATstack){\Gets}\EITstack;$

\subsection{Program control of preemption, and supervisor calls}

So far we have modelled the hardware mechanisms that interleave user
code and interrupt code. The ARM platform additionally provides
mechanisms to do supervisor calls ({\em SVC}), both synchronous and
asynchronous.  
SVC's are
treated as  program-initiated interrupts that are triggered by software
calls to specific platform functions.
Their effect is to switch the execution to specific \OS-provided SVC-handler code.
Asynchronous SVC is typically used to control \OS\ code preemption to
avoid re-entrance (because interrupt handlers would delay a call to the
scheduler if the interrupted task is in \OS\ code). Synchronous SVC is
typically used for kernel calls on platforms supporting dual-mode. In
\theeChronosOS, where \OS\ calls are just function calls, synchronous
SVC is used for direct yielding from application.
Here we present our model of the effect of these
additional platform functions within the framework we introduced above.
We assume code $\SSS$ (resp. $\AAA$) for the
synchronous (resp. asynchronous) SVC handler. 
A \emph{synchronous} SVC is triggered by a call
(in user code)
 to a hardware API function $SVC\_now()$. The effect
of this function is to switch to the execution of
$\SSS$. As with interrupts, the hardware will (atomically)
save context onto the necessary stacks, and set $AT$ to the identifier of the
synchronous SVC task, noted~$SVC_s$.

\noindent
\hspace*{1cm} $SVC\_now() \ \ \equiv \ \ \langle\ \  \ATstack{\Gets}AT+\ATstack;\ 
                                                      EITstack{\Gets}\EIT+\EITstack;\
                                                      AT{\Gets}SVC_s\ \ \rangle $ \hfill (3)

\noindent
The $\langle \cdot \rangle$ notation models the atomic execution of
the instructions, where the atomicity is here ensured by a hardware-enforced atomic mechanism.
The $SVC_s$ task is then modelled as running in
parallel to user code and interrupt code, represented by (c) in~\autoref{fig:wholesys}, with its code wrapped in
\Await-statements using our $\control$ mechanism, and followed
by a return from interrupt (restoring the stacks).

The code for the asynchronous SVC task is modelled the same way, (d)
in~\autoref{fig:wholesys}, but the trigger is delayed. The hardware
provides a function to request an asynchronous supervisor call,
$SVC_a\_Request()$, whose effect is simply to set a bit $SVC_aReq$ to
$True$:

\noindent
\hspace*{.4cm} $SVC_a\_Request() \ \ \equiv \ \ SVC_aReq{\Gets}True $ \hfill (4) 

\noindent
Then at some point in the future 
where
this bit is set, and the $SVC_a$ task is allowed to run (i.e. it is
not masked, and it is allowed to interrupt the running task according
to the interrupt policy),  execution will switch to running
$\AAA$. We model this by having a separate task, (e) in~\autoref{fig:wholesys}, running completely
unguarded, constantly checking if an asynchronous supervisor call has
been requested, and is allowed to run. If it is the case, it resets the
$SVC_aReq$ bit, saves the stacks, and switches to the $SVC_a$ task:

\vspace{.1cm}

\noindent
\hspace*{.4cm} $\ITakeSVCa \ \ \equiv \ \ $ \\
\hspace*{1cm}     $AWAIT\ SVC_aReq \wedge SVC_a \in \EIT-\AT-\ATstack\ \wedge\ (interrupt\_policy\ AT\ SVC_a)\ THEN$\\
\hspace*{1.4cm}        $SVC_aReq{\Gets}False;$\\
\hspace*{1.4cm}        $\ATstack{\Gets}AT+\ATstack;$\\
\hspace*{1.4cm}        $\EITstack{\Gets}\EIT+\EITstack;$\\
\hspace*{1.4cm}        $AT{\Gets}SVC_a;$\\
\hspace*{1cm}     $END$

\noindent
The introduction of these software-triggered interrupts requires
modifying our modelling of return from interrupt. Recall that in
reality the hardware checks for pending interrupts, but in our model
we don't need to model this, since we allow interrupt handlers to run at
any time. However, in the case of the software-triggered
$SVC_a$ interrupt, we need explicitly to model that, on return from
interrupt $I$, the hardware checks whether $SVC_aReq$ is set, and
whether $SVC_a$ is allowed to run (it may have been manually removed
from the $\EIT$ set). To know if $SVC_a$ is allowed to run, we need to
inspect the heads of $\ATstack$ and $\EITstack$ as these are the
context of the task that was interrupted by $I$.
The $\IRet$ function therefore becomes:

\noindent
\hspace*{.4cm} $\IRet(X) \ \ \equiv \ \ $\\
\hspace*{1.4cm}       $IF\ SVC_aReq \wedge SVC_a \in (hd\ \EITstack)-\AT-\ATstack \wedge 
                                    (interrupt\_policy\ (hd\ \ATstack)\ SVC_a)$\\
\hspace*{1.4cm}       $THEN$ $\ \ $
                        $\EIT{\Gets}hd\ \EITstack;$ \\
\hspace*{2.9cm  }         $AT{\Gets}SVC_a;$ \\
\hspace*{1.4cm}       $ELSE$ $\ \ $
                        $(\AT+\ATstack){\Gets}\ATstack;$\\
\hspace*{2.7cm}         $(\EIT+\ATstack){\Gets}\EITstack;$

Our formalisation of the hardware interface is given by the functions
(1)-(4), available to the \OS\ to control interleaving. Additionally,
the functions $\ITake$, $\IRet$ and $\ITakeSVCa$, together with our
$\control$ pre-processing, as presented in~\autoref{fig:wholesys}, form our formal concurrency framework, to
be instantiated to a specific \OS\ by defining $\UU$, $\II$, $\AAA$, and
$\SSS$.

\section{Discussing an Instantiation to a Model of \theeChronosOS}
\label{sec:eChronos model}

For wider usability, we have so far presented a general controlled
  concurrency framework and formal model of an API of hardware
  mechanisms. We have instantiated this general framework to define a
  model of the \eChronos \OS\ scheduling behavior, where tasks are
  allocated priorities by the user, and the scheduler is in charge of
  enforcing that tasks are scheduled according to their priorities,
  i.e. its main functional property is that \emph{whenever application
    code is executing, then it is the highest priority task.} One of
  our longer-term goals is to allow formal verification of such
  correctness properties. Our other aim is to validate our model against
  the real implementation by formal means (formal proof of
  refinement). Validating the hardware abstractions still requires
  informal arguments, though.  Here we discuss the modelling, its
  accuracy and its usability. 

Given the framework described in the previous section, we need to
instantiate $\UU$, $\II$, $\SSS$, and $\AAA$. The instantiation is
given in~\autoref{sec:appendixA}, and the whole Isabelle/HOL model
in~\autoref{sec:appendixB}.

In creating this model there were several issues that we had to
consider to convince ourselves, and more importantly \theeChronosOS
developers, that our model represents reality. The first was that the
way in which we constrain the \OG\ interleaving is accurate. In
particular, we had to investigate when interrupts can occur and what
the hardware does during an interrupt entry and return. 
We also had to ensure that anything we modelled as being atomic
actually is atomic in reality. For the most part this involves the
functionality provided by the hardware that we model, such as the
$\ITake$ and $\IRet$ functions seen in ~\autoref{ssec:interrupt api}.
We believe that we have correctly captured the hardware interrupt
behavior and atomicity, according to the ARM manual~\cite{ARM_manual}.
Additionally, we have been careful {\em not} to use \OG's 
$\langle \cdot \rangle$ atomic statement outside of the hardware
interface. This way, the only remaining assumed atomicity is the one
of single \OG\ statements, which we will need to validate by refinement
proofs when moving on to verification of our model.

Another important issue was the distinction between variables that are
part of \theeChronosOS and the pseudo-variables for hardware
mechanisms. Care was needed to ensure that these hardware
variables are only read and/or modified where allowed to, namely in
the hardware API. Since we target devices with no memory
protection, this requirement will have to be validated for
\theeChronosOS code and will remain an assumption for any user-provided
applications (and could be checked using static analysis).

To justify our use of OG reasoning and to demonstrate that the
mechanisation provided by Isabelle is sufficient to deal with
scalability for system like \theeChronosOS, we have begun initial
verification of our model. As expected, at first there are a very
large number of verification conditions: on the order of
10,000. However, by just defining a method that automatically removes
any redundant conditions we can easily reduce this to under 500.  The
majority of these are then trivial enough to be automatically solved
by standard Isabelle/HOL methods, with the final 10 conditions
requiring human guidance. We believe that this number could be reduced
even further by small improvements to the automation.

\section{Related work and Conclusions}
Frameworks for reasoning about shared variable programs have been
around for more than 30 years. \OG\ was the first one to be
proposed~\cite{Owicki_Gries_76}; much derived work has been done since, addressing specific
requirements (compositionality~\cite{Jones_83}, resource
separation~\cite{OHearn_07}, etc). These frameworks have mainly been
used to prove the correctness of concurrency algorithms or
protocols. Here we target low-level high-performance \OS\ code.
Similarily, higher-level conceptual tools such as
monitors~\cite{Hoare_74} and conditional critical
regions~\cite{Hansen_72} decrease the proof burden, but impose a
performance penalty, a trade-off usually worth making for clarity,
except for minimal high-speed \OS-kernel application where efficiency
is crucial.

Formal verification of operating systems, kernels, and hypervisors has
been the focus of important recent research (for which see~\cite{Klein_09} for
an overview). Successfully verified systems generally either run on
uniprocessor platforms with interrupts mostly disabled
(e.g.~\cite{Klein_AEMSKH_14, Yang_Hawblitzel_10, Richards_10}), or
their verification does not take interrupts into account
(e.g.~\cite{Huang_11}).
In~\cite{Feng_SGD_09}, a Hoare-logic-based framework is proposed to
certify low-level system code involving interrupts and preemptive
tasks, but the scheduler and context switching tasks are still executed
with interrupts disabled, and interrupt handlers cannot be
interrupted. In contrast, our work supports nested interrupts and a
preemptible scheduler.
A proof of correctness of the FreeRTOS scheduler is proposed
in~\cite{Ferreira_HQ_12}; the proof does not include the context
switch itself and focuses on the scheduler policy (picking the next
task). This is complementary to our work, where we leave the policy
generic and assume it will pick the highest priority task.

To our knowledge, our extended OG framework with controlled
concurrency is the first to support reasoning about low-level system
code that is fully preemptible, including scheduler code, with support
for nested interrupts. We have successfully instantiated it to
formalise the scheduling behavior of \theeChronosOS, a real-world,
deployed, embedded \OS. Our promising initial verification work
indicates that we will be able to formally prove important functional
properties involving complex concurrency reasoning about highly shared
low-level variables.

\nocite{*}
\bibliographystyle{eptcs}

\newpage

\appendix
\section{Instantiation of our Controlled OG Framework to {the \textit{eChronos}~OS}}
\label{sec:appendixA}

Instantiating the framework described in~\autoref{sec:api} to
\theeChronosOS requires to instantiate $\UU$, $\II$, $\SSS$, and
$\AAA$. The instantiation is as follows.

\vspace{.1cm}

\noindent
\hspace*{.4cm}  $\II\ \ \equiv\ \ E{\Gets}change\_events{\DSemi}$\\
\hspace*{1.6cm}                    $SVC_a\_Request(){\DSemi}$\\[.15cm]
\hspace*{.4cm}  $\UU_j\ \ \equiv\ \ syscall\_block\ \ \equiv\ \ SVC_aDisable(){\DSemi}$\\
\hspace*{4.65cm}                   $R{\Gets} R(j:=False){\DSemi}$\\
\hspace*{4.65cm}                   $SVC\_now(){\DSemi} $\\
\hspace*{4.65cm}                   $SVC_aEnable(){\DSemi}$\\
\hspace*{4.65cm}                   $\While \neg SVC_aReq\ \Do\ \ SKIP\ \ \End{\DSemi}$\\[.15cm]
\hspace*{.4cm}  $\SSS\ \ \equiv \ schedule{\DSemi}$\\
\hspace*{1.6cm}            $context\_switch \ True{\DSemi}$\\[.15cm]
\hspace*{.4cm}  $\AAA\ \ \equiv \ schedule{\DSemi}$\\
\hspace*{1.6cm}            $context\_switch \ False{\DSemi}$

\vspace{.1cm}

Since we are focusing on the scheduling behavior, we only model the
parts that may influence the scheduling decisions, i.e.\ deciding which
task should be the next to run. These decisions depend on (i) which
are the runnable tasks, and (ii) the set of events signaled by
interrupt handlers, which may influence which tasks are runnable. We
use the variable $\R$ for the mapping from task identifier to a Boolean
value indicating whether the task is runnable, and the variable $\E$ for
the set of events.

The interrupt code $\II$ is mainly a user-provided interrupt handler,
which is only allowed to call one specific \OS\ function to change the
set of events. Since this might change which tasks are runnable the
scheduler needs to run to update the set of runnable tasks and
potentially switch to a higher priority task. To avoid re-entrant \OS\
code, the interrupt handler only flags the need for the scheduler to
run (by requesting an asynchronous system call). This request must be
handled before application code is run again. 
The function $change\_events$ represents a non-deterministic
update. The rest of the interrupt handler's functionality is not
represented, as it should not be relevant to the scheduling behavior.

$\SSS$ and $\AAA$ are almost identical and represent the scheduler
code. The main job of the scheduler is to pick a new task to run, by
first updating the runnable mapping $\R$ taking into account the set
of unprocessed events $\E$, and then picking the task to run according
to the scheduling policy in place. Once the task is chosen, a context
switch is performed, storing the old task and placing the new task on
the stack. The full
details of how $schedule$ and $context\_switch$ are modelled can be
found in \autoref{ssec:appendix program}.

Finally, the majority of \theeChronosOS code is in $\UU$, which
represents application code (kept generic here) and calls to any of
the \OS\ API functions. 
We model
application code only as potentially making an \OS\ call, and we only
model a single call that is representative of how the variables that
we are interested in can be modified.
The block syscall modifies $R$ so that task $U_j$ is not runnable and
then yields by invoking $\SVCs$ via $SVC\_now$. To ensure that it is
not re-entrant, the \OS\ call is wrapped between disable and enable
functions for the $\SVCa$ interrupt and is followed by a loop waiting
for $SVC_aReq$ to be set to $False$. As $\SVCa$ is the only routine
that sets $SVC_aReq$ to $False$, this ensures that, if required,
$\SVCa$ executes before control is returned to the user application.
The functions $SVC_aDisable()$ and $SVC_aEnable()$ are defined as follows.

\vspace*{.2cm}

\noindent
\hspace*{.4cm} $SVC_aDisable() \ \ \equiv \ \ \EIT{\Gets}\EIT-{SVC_a} $\\
\hspace*{.4cm} $SVC_aEnable() \ \ \equiv \ \ \EIT{\Gets}\EIT\cup{SVC_a} $

\newpage
\section{Formal model of the eChronos \OS\ scheduling behaviour in Isabelle}
\label{sec:appendixB}
\begin{isabellebody}%
\setisabellecontext{echronos}%
\isadelimtheory
\endisadelimtheory
\isatagtheory
\endisatagtheory
{\isafoldtheory}%
\isadelimtheory
\endisadelimtheory
\begin{isamarkuptext}%
We present here a model of the ARM Cortex-M4 version of the eChronos
OS scheduling behaviour, formalised in Isabelle/HOL. It is based on
Leonor Prensa's formalisation of Owicki-Gries in Isabelle/HOL.%
\end{isamarkuptext}%
\isamarkuptrue%
\isamarkupsubsection{State%
}
\isamarkuptrue%
\begin{isamarkuptext}%
A routine is just a natural number; we add routines for both the
$SVC_s$ handler and the $SVC_a$ handler, user routines have numbers
from \isa{{\isadigit{2}}} to \isa{nbUsers{\isacharplus}{\isadigit{2}}}(excluded) and interrupt routines
have numbers from \isa{nbUsers{\isacharplus}{\isadigit{2}}} to \isa{nbUsers{\isacharplus}nbInts{\isacharplus}{\isadigit{2}}}
(excluded). The first user to run is arbitrarily chosen to be the
first one.%
\end{isamarkuptext}%
\isamarkuptrue%
\isacommand{type{\isacharunderscore}synonym}\isamarkupfalse%
\ routine\ {\isacharequal}\ nat\isanewline
\isanewline
\isacommand{consts}\isamarkupfalse%
\ nbUsers\ {\isacharcolon}{\isacharcolon}\ nat\isanewline
\isacommand{consts}\isamarkupfalse%
\ nbInts\ {\isacharcolon}{\isacharcolon}\ nat\isanewline
\isanewline
\isacommand{abbreviation}\isamarkupfalse%
\ {\isachardoublequoteopen}nbRoutines\ {\isasymequiv}\ nbUsers{\isacharplus}nbInts{\isachardoublequoteclose}\isanewline
\isanewline
\isacommand{abbreviation}\isamarkupfalse%
\ {\isachardoublequoteopen}SVC\isactrlsub s\ {\isasymequiv}\ {\isadigit{0}}{\isachardoublequoteclose}\isanewline
\isacommand{abbreviation}\isamarkupfalse%
\ {\isachardoublequoteopen}SVC\isactrlsub a\ {\isasymequiv}\ {\isadigit{1}}{\isachardoublequoteclose}\isanewline
\isanewline
\isacommand{definition}\isamarkupfalse%
\ {\isachardoublequoteopen}user{\isadigit{0}}\ {\isasymequiv}\ {\isadigit{2}}{\isachardoublequoteclose}\isanewline
\isacommand{definition}\isamarkupfalse%
\ {\isachardoublequoteopen}U\ {\isasymequiv}\ {\isacharbraceleft}user{\isadigit{0}}{\isachardot}{\isachardot}{\isacharless}user{\isadigit{0}}\ {\isacharplus}\ nbUsers{\isacharbraceright}{\isachardoublequoteclose}\isanewline
\isacommand{definition}\isamarkupfalse%
\ {\isachardoublequoteopen}I\ {\isasymequiv}\ {\isacharbraceleft}user{\isadigit{0}}\ {\isacharplus}\ nbUsers{\isachardot}{\isachardot}{\isacharless}user{\isadigit{0}}\ {\isacharplus}\ nbUsers\ {\isacharplus}\ nbInts{\isacharbraceright}{\isachardoublequoteclose}\isanewline
\isacommand{definition}\isamarkupfalse%
\ {\isachardoublequoteopen}I{\isacharprime}\ {\isasymequiv}\ I\ {\isasymunion}\ {\isacharbraceleft}SVC\isactrlsub a{\isacharbraceright}{\isachardoublequoteclose}%
\begin{isamarkuptext}%
\noindent
A state is composed of all the hardware variables plus the program variables that
the targeted invariant or property relies on.%
\end{isamarkuptext}%
\isamarkuptrue%
\isacommand{record}\isamarkupfalse%
\ {\isacharprime}a\ state\ {\isacharequal}\isanewline
\ \ EIT\ {\isacharcolon}{\isacharcolon}\ {\isachardoublequoteopen}routine\ set{\isachardoublequoteclose}\ \ \ \ \ \ \ \ \ \ \ \ \ \ \ \ \ \ \ \ \ \ \ \ \ \ \ \ \ \ \ \ \ \ \ \ \ \ \ \ \ \ \ \ \ \ \ \ \ \ \ \ \ \ \ \ %
\isamarkupcmt{the set of enabled interrupt tasks%
}
\isanewline
\ \ SVC\isactrlsub aReq\ {\isacharcolon}{\isacharcolon}\ bool\ \ \ \ \ \ \ \ \ \ \ \ \ \ \ \ \ \ \ \ \ \ \ \ \ \ \ \ \ \ \ \ \ \ \ \ \ \ \ \ \ \ \ \ \ \ \ \ \ \ \ \ \ \ \ \ \ %
\isamarkupcmt{the $SVC_a$ requested bit%
}
\isanewline
\ \ AT\ {\isacharcolon}{\isacharcolon}\ routine\ \ \ \ \ \ \ \ \ \ \ \ \ \ \ \ \ \ \ \ \ \ \ \ \ \ \ \ \ \ \ \ \ \ \ \ \ \ \ \ \ \ \ \ \ \ \ \ \ \ \ \ \ \ \ \ \ \ \ \ \ \ \ %
\isamarkupcmt{the active routine%
}
\isanewline
\ \ ATStack\ {\isacharcolon}{\isacharcolon}\ {\isachardoublequoteopen}routine\ list{\isachardoublequoteclose}\ \ \ \ \ \ \ \ \ \ \ \ \ \ \ \ \ \ \ \ \ \ \ \ \ \ \ \ \ \ \ \ \ \ \ \ \ \ \ \ \ \ \ \ \ \ \ \ \ %
\isamarkupcmt{the stack of suspended routines%
}
\isanewline
\isanewline
\ \ curUser\ {\isacharcolon}{\isacharcolon}\ routine\ \ \ \ \ \ \ \ \ \ \ \ \ \ \ \ \ \ \ \ \ \ \ \ \ \ \ \ \ \ \ \ \ \ \ \ \ \ \ \ \ \ \ \ \ \ \ \ \ \ \ \ \ \ \ %
\isamarkupcmt{current user task%
}
\isanewline
\ \ contexts\ {\isacharcolon}{\isacharcolon}\ {\isachardoublequoteopen}routine\ {\isasymRightarrow}\ {\isacharparenleft}bool\ {\isasymtimes}\ routine\ list{\isacharparenright}\ option{\isachardoublequoteclose}\ \ \ \ \ %
\isamarkupcmt{stored contexts%
}
\isanewline
\ \ R\ {\isacharcolon}{\isacharcolon}\ {\isachardoublequoteopen}routine\ {\isasymRightarrow}\ bool\ option{\isachardoublequoteclose}\ \ \ \ \ \ \ \ \ \ \ \ \ \ \ \ \ \ \ \ \ \ \ \ \ \ \ \ \ \ \ \ \ \ \ \ \ \ \ \ \ %
\isamarkupcmt{Runnable threads%
}
\isanewline
\ \ E\ {\isacharcolon}{\isacharcolon}\ {\isachardoublequoteopen}nat\ set{\isachardoublequoteclose}\ \ \ \ \ \ \ \ \ \ \ \ \ \ \ \ \ \ \ \ \ \ \ \ \ \ \ \ \ \ \ \ \ \ \ \ \ \ \ \ \ \ \ \ \ \ \ \ \ \ \ \ \ \ \ \ \ \ \ \ \ \ \ \ \ \ %
\isamarkupcmt{Events set (current)%
}
\isanewline
\ \ E{\isacharunderscore}tmp\ {\isacharcolon}{\isacharcolon}\ {\isachardoublequoteopen}nat\ set{\isachardoublequoteclose}\ \ \ \ \ \ \ \ \ \ \ \ \ \ \ \ \ \ \ \ \ \ \ \ \ \ \ \ \ \ \ \ \ \ \ \ \ \ \ \ \ \ \ \ \ \ \ \ \ \ \ \ \ \ \ \ \ \ \ %
\isamarkupcmt{Temporary events set%
}
\isanewline
\ \ nextT\ {\isacharcolon}{\isacharcolon}\ {\isachardoublequoteopen}routine\ option{\isachardoublequoteclose}\ \ \ \ \ \ \ \ \ \ \ \ \ \ \ \ \ \ \ \ \ \ \ \ \ \ \ \ \ \ \ \ \ \ \ \ \ \ \ \ \ \ \ \ \ \ \ \ %
\isamarkupcmt{the next Task%
}
\isamarkupsubsection{Controlled Owicki-Gries reasoning%
}
\isamarkuptrue%
\begin{isamarkuptext}%
The model of parallel composition allows more interleaving than the
real execution, where only enabled routines can run. To model this we
extend the OG formalisation with our controlled concurrency mechanism;
we use the \isa{AT} variable and wrap every instruction of routine
\isa{r} in an \isa{AWAIT\ {\isasymlbrace}AT\ {\isacharequal}\ r{\isasymrbrace}} statement. The function
\isa{add{\isacharunderscore}await{\isacharunderscore}bare{\isacharunderscore}com} performs this process. It recursively
traverses the command tree, using the given property to construct the
\isa{AWAIT} statement which is added as required. The full
definition of \isa{add{\isacharunderscore}await{\isacharunderscore}bare{\isacharunderscore}com} is not shown here.%
\end{isamarkuptext}%
\isamarkuptrue%
\isacommand{definition}\isamarkupfalse%
\ control\isanewline
\isakeyword{where}\isanewline
\ \ {\isachardoublequoteopen}control\ r\ c\ {\isasymequiv}\ add{\isacharunderscore}await{\isacharunderscore}bare{\isacharunderscore}com\ {\isasymlbrace}AT\ {\isacharequal}\ r{\isasymrbrace}\ c{\isachardoublequoteclose}%
\isamarkupsubsection{Generic scheduling policy, handling of events and interrupt policy%
}
\isamarkuptrue%
\begin{isamarkuptext}%
The scheduling policy (picking the next thread, given the list of
runnable threads) is left unspecified here; as well as the updating of
this runnable list, given a list of events. The interrupt policy
(which interrupts are allowed to run, given the currently running
routine) is also left unspecified.%
\end{isamarkuptext}%
\isamarkuptrue%
\isacommand{consts}\isamarkupfalse%
\ sched{\isacharunderscore}policy\ {\isacharcolon}{\isacharcolon}\ {\isachardoublequoteopen}{\isacharparenleft}routine\ {\isasymRightarrow}\ bool\ option{\isacharparenright}\ {\isasymRightarrow}\ routine\ option{\isachardoublequoteclose}\isanewline
\isacommand{consts}\isamarkupfalse%
\ handle{\isacharunderscore}events\ {\isacharcolon}{\isacharcolon}\ {\isachardoublequoteopen}nat\ set\ {\isasymRightarrow}\ {\isacharparenleft}routine\ {\isasymRightarrow}\ bool\ option{\isacharparenright}\ {\isasymRightarrow}\ routine\ {\isasymRightarrow}\ bool\ option{\isachardoublequoteclose}\isanewline
\isacommand{consts}\isamarkupfalse%
\ interrupt{\isacharunderscore}policy\ {\isacharcolon}{\isacharcolon}\ {\isachardoublequoteopen}routine\ {\isasymRightarrow}\ routine\ set{\isachardoublequoteclose}%
\isamarkupsubsection{A model of  hardware interface%
}
\isamarkuptrue%
\begin{isamarkuptext}%
The following two functions are used to enable and disable the
$\mathrm{SVC_a}$ interrupt. They do this by either adding or
removing $SVC_a$ from the $EIT$ set.%
\end{isamarkuptext}%
\isamarkuptrue%
\isacommand{definition}\isamarkupfalse%
\isanewline
\ \ SVC\isactrlsub aEnable\isanewline
\isakeyword{where}\isanewline
\ \ {\isachardoublequoteopen}SVC\isactrlsub aEnable\ {\isasymequiv}\ EIT\ {\isacharcolon}{\isacharequal}\ EIT\ {\isasymunion}\ {\isacharbraceleft}SVC\isactrlsub a{\isacharbraceright}{\isachardoublequoteclose}\isanewline
\isanewline
\isacommand{definition}\isamarkupfalse%
\isanewline
\ \ SVC\isactrlsub aDisable\isanewline
\isakeyword{where}\isanewline
\ \ {\isachardoublequoteopen}SVC\isactrlsub aDisable\ {\isasymequiv}\ EIT\ {\isacharcolon}{\isacharequal}\ EIT\ {\isacharminus}\ {\isacharbraceleft}SVC\isactrlsub a{\isacharbraceright}{\isachardoublequoteclose}%
\begin{isamarkuptext}%
\noindent

$\mathitbf{ITake}\ i$ models the hardware mechanism that traps to the
handler for interrupt $i$. First it checks whether the interrupt is
enabled, is not already being handled and is a higher priority than
the current routine. When these conditions are satisfied then the
context\footnote{We only model the part of the context relevant to
controlling the interleaving. Here that is just the previous value of
$AT$, which can be thought of as corresponding to the program counter.
Note that this is in contrast to the model from
\autoref{ssec:interrupt api}, which also stores the value of $EIT$.
This is because ARM does not save the mask status when an interrupt
occurs, and it is up to the interrupt handlers to ensure that the
interrupt mask is preserved.} of the previous task is saved on a stack
and $AT$ is set to $i$.%
\end{isamarkuptext}%
\isamarkuptrue%
\isacommand{definition}\isamarkupfalse%
\isanewline
\ \ ITake\isanewline
\isakeyword{where}\isanewline
\ \ {\isachardoublequoteopen}ITake\ i\ {\isasymequiv}\isanewline
\ \ \ \ AWAIT\ i\ {\isasymin}\ EIT\ {\isacharminus}\ {\isacharbraceleft}AT{\isacharbraceright}\ {\isacharminus}\ set\ ATStack\ {\isasymand}\ i\ {\isasymin}\ interrupt{\isacharunderscore}policy\ AT\isanewline
\ \ \ \ THEN\isanewline
\ \ \ \ \ {\isasymlangle}ATStack\ {\isacharcolon}{\isacharequal}\ AT\ {\isacharhash}\ ATStack{\isacharcomma}{\isacharcomma}\ AT\ {\isacharcolon}{\isacharequal}\ i{\isasymrangle}\isanewline
\ \ \ \ END{\isachardoublequoteclose}%
\begin{isamarkuptext}%
\noindent
Similarly to above, $\mathitbf{SVC_aTake}$ models the hardware
mechanism that traps to the $\mathrm{SVC_a}$ handler. It is almost
exactly the same as $\mathitbf{ITake}\ i$, but because we can observe
when $\mathrm{SVC_a}$ is requested we now also require that $SVC_aReq$
is True before it can begin executing. It also sets $SVC_aReq$ to
False while setting $AT$ to $\mathrm{SVC_a}$.%
\end{isamarkuptext}%
\isamarkuptrue%
\isacommand{definition}\isamarkupfalse%
\isanewline
\ \ SVC\isactrlsub aTake\isanewline
\isakeyword{where}\isanewline
\ \ {\isachardoublequoteopen}SVC\isactrlsub aTake\ {\isasymequiv}\isanewline
\ \ \ \ AWAIT\ SVC\isactrlsub aReq\ {\isasymand}\ SVC\isactrlsub a\ {\isasymin}\ EIT\ {\isacharminus}\ {\isacharbraceleft}AT{\isacharbraceright}\ {\isacharminus}\ set\ ATStack\ {\isasymand}\ SVC\isactrlsub a\ {\isasymin}\ interrupt{\isacharunderscore}policy\ AT\isanewline
\ \ \ \ THEN\isanewline
\ \ \ \ \ \ {\isasymlangle}ATStack\ {\isacharcolon}{\isacharequal}\ AT\ {\isacharhash}\ ATStack{\isacharcomma}{\isacharcomma}\isanewline
\ \ \ \ \ \ \ AT\ {\isacharcolon}{\isacharequal}\ SVC\isactrlsub a{\isacharcomma}{\isacharcomma}\ SVC\isactrlsub aReq\ {\isacharcolon}{\isacharequal}\ False{\isasymrangle}\isanewline
\ \ \ \ END{\isachardoublequoteclose}%
\begin{isamarkuptext}%
\noindent
$\mathitbf{IRet}$ models the hardware mechanism used to return from an
interrupt. The main action it performs is to restore the context of
the interrupted routine. It does this by setting $AT$ to the head
$ATStack$, which is then removed from the stack. However, if there is
a pending interrupt that is now allowed to run then $\mathitbf{IRet}$
will transfer control directly to this interrupt instead of restoring
the stored context. Due to the construction of our model we only need
to ensure that this happens for $\mathrm{SVC_a}$, as it is the only
interrupt that we can observe has occured.%
\end{isamarkuptext}%
\isamarkuptrue%
\isacommand{definition}\isamarkupfalse%
\isanewline
\ \ IRet\isanewline
\isakeyword{where}\isanewline
\ \ {\isachardoublequoteopen}IRet\ {\isasymequiv}\isanewline
\ \ \ \ {\isasymlangle}IF\ SVC\isactrlsub aReq\ {\isasymand}\ SVC\isactrlsub a\ {\isasymin}\ EIT\ {\isacharminus}\ set\ ATStack\ {\isasymand}\ SVC\isactrlsub a\ {\isasymin}\ interrupt{\isacharunderscore}policy\ {\isacharparenleft}hd\ ATStack{\isacharparenright}\isanewline
\ \ \ \ \ \ THEN\ AT\ {\isacharcolon}{\isacharequal}\ SVC\isactrlsub a{\isacharcomma}{\isacharcomma}\ SVC\isactrlsub aReq\ {\isacharcolon}{\isacharequal}\ False\isanewline
\ \ \ \ \ \ ELSE\ AT\ {\isacharcolon}{\isacharequal}\ hd\ ATStack{\isacharcomma}{\isacharcomma}\ ATStack\ {\isacharcolon}{\isacharequal}\ tl\ ATStack\isanewline
\ \ \ \ \ FI{\isasymrangle}{\isachardoublequoteclose}%
\begin{isamarkuptext}%
\noindent
When $SVC\_now$ is called it triggers an $\mathrm{SVC_s}$
interrupt to occur, which is then immediately handled. The effect of
this function is similar to that of $\mathitbf{ITake}$, the active
task is saved on the stack and then $AT$ is set to $SVC_s$. We
implicitly assume that $\mathrm{SVC_s}$ is enabled when
$SVC\_now$ is called, as if this was not true in reality then the
hardware would trigger an abort exception.%
\end{isamarkuptext}%
\isamarkuptrue%
\isacommand{definition}\isamarkupfalse%
\isanewline
\ \ SVC{\isacharunderscore}now\isanewline
\isakeyword{where}\isanewline
\ \ {\isachardoublequoteopen}SVC{\isacharunderscore}now\ {\isasymequiv}\ {\isasymlangle}ATStack\ {\isacharcolon}{\isacharequal}\ AT\ {\isacharhash}\ ATStack{\isacharcomma}{\isacharcomma}\ AT\ {\isacharcolon}{\isacharequal}\ SVC\isactrlsub s{\isasymrangle}{\isachardoublequoteclose}%
\begin{isamarkuptext}%
\noindent
$SVC_aRequest$ is used to request that $\mathrm{SVC_s}$
occurs as soon as it is next possible.%
\end{isamarkuptext}%
\isamarkuptrue%
\isacommand{definition}\isamarkupfalse%
\isanewline
\ \ SVC\isactrlsub aRequest\isanewline
\isakeyword{where}\isanewline
\ \ {\isachardoublequoteopen}SVC\isactrlsub aRequest\ {\isasymequiv}\ SVC\isactrlsub aReq\ {\isacharcolon}{\isacharequal}\ True{\isachardoublequoteclose}%
\isamarkupsubsection{Model of the eChronos OS%
}
\isamarkuptrue%
\begin{isamarkuptext}%
\label{ssec:appendix program}%
\end{isamarkuptext}%
\isamarkuptrue%
\begin{isamarkuptext}%
The eChronos OS uses $SVC_s$ and $SVC_a$ interrupt handlers to
implement scheduling. The scheduler function chooses a new task to run
by first updating the runnable mapping $R$ before using
whichever scheduler policy is in place to pick a task from among the
runnable ones. To update the runnable mapping, the function
$handle\_events$ is used, with this function being left
non-deterministic. After the execution of this function, the variable
$E$ needs to be cleared to indicate that the events have been
processed. However, the scheduler may itself be interrupted. If an
interrupt occurs between the execution of $handle\_events$ and the
reset of $E$, the interrupt handler might have modified
$E$ with new events to be processed (and flagged a request
for the scheduler to run). On return from interrupt, because the
scheduler is itself an interrupt and is not re-entrant, its execution
resumes, and so $E$ should not be cleared. Instead we save
its value before running $handle\_events$, and only remove those saved
events that have indeed been processed. When the scheduler will
return, the hardware will check if there are still any pending
requests for the scheduler to run, and re-run it if required.%
\end{isamarkuptext}%
\isamarkuptrue%
\isacommand{definition}\isamarkupfalse%
\isanewline
\ \ schedule\isanewline
\isakeyword{where}\isanewline
\ \ {\isachardoublequoteopen}schedule\ {\isasymequiv}\isanewline
\ \ \ \ nextT\ {\isacharcolon}{\isacharequal}\ None{\isacharsemicolon}{\isacharsemicolon}\isanewline
\ \ \ \ WHILE\ nextT\ {\isacharequal}\ None\isanewline
\ \ \ \ DO\isanewline
\ \ \ \ \ \ E{\isacharunderscore}tmp\ {\isacharcolon}{\isacharequal}\ E{\isacharsemicolon}{\isacharsemicolon}\isanewline
\ \ \ \ \ \ R\ {\isacharcolon}{\isacharequal}\ handle{\isacharunderscore}events\ E{\isacharunderscore}tmp\ R{\isacharsemicolon}{\isacharsemicolon}\isanewline
\ \ \ \ \ \ E\ {\isacharcolon}{\isacharequal}\ E\ {\isacharminus}\ E{\isacharunderscore}tmp{\isacharsemicolon}{\isacharsemicolon}\isanewline
\ \ \ \ \ \ nextT\ {\isacharcolon}{\isacharequal}\ sched{\isacharunderscore}policy{\isacharparenleft}R{\isacharparenright}\isanewline
\ \ \ \ OD{\isachardoublequoteclose}%
\begin{isamarkuptext}%
\noindent
Once the schedule functions has executed, the $context\_switch$
function is called. This function, as the name suggests, saves the
context of the old user task that was previously on the hardware
stack, and then replaces it with the context of the task chosen by the
scheduler. To do this the function first stores whether the previous
user task had $SVC_a$ enabled,\footnote{This is identified by the
boolean passed to context\_switch. If $context\_switch$ is being
called by $SVC_a$ then clearly $SVC_a$ was previously enabled, while
by design we know that $SVC_s$ is only called when $SVC_a$ is
disabled.} along with the current value of $ATStack$. It then loads
the stack that existed when the new task was last executing. Lastly,
$SVC_a$ is enabled or disabled as required by the new task.%
\end{isamarkuptext}%
\isamarkuptrue%
\isacommand{definition}\isamarkupfalse%
\isanewline
\ \ context{\isacharunderscore}switch\isanewline
\isakeyword{where}\isanewline
\ \ {\isachardoublequoteopen}context{\isacharunderscore}switch\ preempt{\isacharunderscore}enabled\ {\isasymequiv}\isanewline
\ \ \ \ contexts\ {\isacharcolon}{\isacharequal}\ contexts\ {\isacharparenleft}curUser\ {\isasymmapsto}\ {\isacharparenleft}preempt{\isacharunderscore}enabled{\isacharcomma}\ ATStack{\isacharparenright}{\isacharparenright}{\isacharsemicolon}{\isacharsemicolon}\isanewline
\ \ \ \ curUser\ {\isacharcolon}{\isacharequal}\ the\ nextT{\isacharsemicolon}{\isacharsemicolon}\isanewline
\ \ \ \ ATStack\ {\isacharcolon}{\isacharequal}\ snd\ {\isacharparenleft}the\ {\isacharparenleft}contexts\ {\isacharparenleft}curUser{\isacharparenright}{\isacharparenright}{\isacharparenright}{\isacharsemicolon}{\isacharsemicolon}\ \isanewline
\ \ \ \ IF\ fst\ {\isacharparenleft}the\ {\isacharparenleft}contexts\ {\isacharparenleft}curUser{\isacharparenright}{\isacharparenright}{\isacharparenright}\isanewline
\ \ \ \ \ THEN\ SVC\isactrlsub aEnable\isanewline
\ \ \ \ \ ELSE\ SVC\isactrlsub aDisable\isanewline
\ \ \ \ FI{\isachardoublequoteclose}%
\begin{isamarkuptext}%
\noindent
Finally, we combine everything to construct the full eChronos OS
model. First, the state is initialised with the correct starting
values. Following this, the various routines are run in parallel, with
concurrency controlled as required through the use of \isa{control}.%
\end{isamarkuptext}%
\isamarkuptrue%
\isacommand{definition}\isamarkupfalse%
\isanewline
\ \ eChronos{\isacharunderscore}OS{\isacharunderscore}model\isanewline
\isakeyword{where}\isanewline
\ \ {\isachardoublequoteopen}eChronos{\isacharunderscore}OS{\isacharunderscore}model\ change{\isacharunderscore}runnables\ change{\isacharunderscore}events\ {\isasymequiv}\isanewline
\ \ {\isacharparenleft}EIT\ {\isacharcolon}{\isacharequal}\ I{\isacharprime}{\isacharcomma}{\isacharcomma}\isanewline
\ \ \ SVC\isactrlsub aReq\ {\isacharcolon}{\isacharequal}\ False{\isacharcomma}{\isacharcomma}\isanewline
\ \ \ AT\ {\isacharcolon}{\isacharequal}\ user{\isadigit{0}}{\isacharcomma}{\isacharcomma}\isanewline
\ \ \ ATStack\ {\isacharcolon}{\isacharequal}\ {\isacharbrackleft}{\isacharbrackright}{\isacharcomma}{\isacharcomma}\isanewline
\ \ \ curUser\ {\isacharcolon}{\isacharequal}\ user{\isadigit{0}}{\isacharcomma}{\isacharcomma}\isanewline
\ \ \ contexts\ {\isacharcolon}{\isacharequal}\ {\isacharparenleft}{\isasymlambda}n{\isachardot}\ if\ n{\isasymin}U\ then\ Some\ {\isacharparenleft}True{\isacharcomma}\ {\isacharbrackleft}n{\isacharbrackright}{\isacharparenright}\ else\ None{\isacharparenright}{\isacharcomma}{\isacharcomma}\isanewline
\ \ \ R\ {\isacharcolon}{\isacharequal}\ {\isacharparenleft}{\isasymlambda}n{\isachardot}\ if\ n{\isasymin}U\ then\ Some\ True\ else\ None{\isacharparenright}{\isacharcomma}{\isacharcomma}\isanewline
\ \ \ E\ {\isacharcolon}{\isacharequal}\ {\isacharbraceleft}{\isacharbraceright}{\isacharcomma}{\isacharcomma}\isanewline
\ \ \ E{\isacharunderscore}tmp\ {\isacharcolon}{\isacharequal}\ {\isacharbraceleft}{\isacharbraceright}{\isacharcomma}{\isacharcomma}\isanewline
\ \ \ nextT\ {\isacharcolon}{\isacharequal}\ None{\isacharcomma}{\isacharcomma}\isanewline
\ \ {\isacharparenleft}COBEGIN\isanewline
\ \ \ \ {\isacharparenleft}{\isacharasterisk}\ SVC\isactrlsub a{\isacharunderscore}take\ {\isacharasterisk}{\isacharparenright}\isanewline
\ \ \ \ WHILE\ True\isanewline
\ \ \ \ DO\isanewline
\ \ \ \ \ \ SVC\isactrlsub aTake\isanewline
\ \ \ \ OD\isanewline
\isanewline
\ \ \ \ {\isasymparallel}\isanewline
\isanewline
\ \ \ \ {\isacharparenleft}{\isacharasterisk}\ SVC\isactrlsub a\ {\isacharasterisk}{\isacharparenright}\isanewline
\ \ \ \ WHILE\ True\isanewline
\ \ \ \ DO\isanewline
\ \ \ \ \ \ {\isacharparenleft}control\ SVC\isactrlsub a\ {\isacharparenleft}\isanewline
\ \ \ \ \ \ schedule{\isacharsemicolon}{\isacharsemicolon}\isanewline
\ \ \ \ \ \ context{\isacharunderscore}switch\ True{\isacharsemicolon}{\isacharsemicolon}\isanewline
\ \ \ \ \ \ IRet{\isacharparenright}{\isacharparenright}\isanewline
\ \ \ \ OD\isanewline
\isanewline
\ \ \ \ {\isasymparallel}\isanewline
\isanewline
\ \ \ \ {\isacharparenleft}{\isacharasterisk}\ SVC\isactrlsub s\ {\isacharasterisk}{\isacharparenright}\isanewline
\ \ \ \ WHILE\ True\isanewline
\ \ \ \ DO\isanewline
\ \ \ \ \ \ {\isacharparenleft}control\ SVC\isactrlsub s\ {\isacharparenleft}\isanewline
\ \ \ \ \ \ schedule{\isacharsemicolon}{\isacharsemicolon}\isanewline
\ \ \ \ \ \ context{\isacharunderscore}switch\ False{\isacharsemicolon}{\isacharsemicolon}\isanewline
\ \ \ \ \ \ IRet{\isacharparenright}{\isacharparenright}\isanewline
\ \ \ \ OD\isanewline
\isanewline
\ \ \ \ {\isasymparallel}\isanewline
\isanewline
\ \ \ \ SCHEME\ {\isacharbrackleft}{\isadigit{0}}\ {\isasymle}\ i\ {\isacharless}\ nbRoutines{\isacharbrackright}\isanewline
\ \ \ \ IF\ {\isacharparenleft}i{\isasymin}I{\isacharparenright}\ THEN\isanewline
\isanewline
\ \ \ \ {\isacharparenleft}{\isacharasterisk}\ Interrupts\ {\isacharasterisk}{\isacharparenright}\isanewline
\ \ \ \ WHILE\ True\isanewline
\ \ \ \ DO\isanewline
\ \ \ \ \ \ ITake\ i{\isacharsemicolon}{\isacharsemicolon}\isanewline
\isanewline
\ \ \ \ \ \ {\isacharparenleft}control\ i\ {\isacharparenleft}\isanewline
\ \ \ \ \ \ E\ {\isacharcolon}{\isacharequal}\ change{\isacharunderscore}events{\isacharsemicolon}{\isacharsemicolon}\isanewline
\ \ \ \ \ \ SVC\isactrlsub aRequest{\isacharsemicolon}{\isacharsemicolon}\isanewline
\isanewline
\ \ \ \ \ \ IRet{\isacharparenright}{\isacharparenright}\isanewline
\ \ \ \ OD\isanewline
\isanewline
\ \ \ \ ELSE\isanewline
\ \ \ \ {\isacharparenleft}{\isacharasterisk}\ Users\ {\isacharasterisk}{\isacharparenright}\isanewline
\ \ \ \ WHILE\ True\isanewline
\ \ \ \ DO\isanewline
\ \ \ \ \ \ {\isacharparenleft}control\ i\ {\isacharparenleft}\isanewline
\ \ \ \ \ \ SVC\isactrlsub aDisable{\isacharsemicolon}{\isacharsemicolon}\isanewline
\ \ \ \ \ \ R\ {\isacharcolon}{\isacharequal}\ R\ {\isacharparenleft}i\ {\isasymmapsto}\ False{\isacharparenright}{\isacharsemicolon}{\isacharsemicolon}\isanewline
\ \ \ \ \ \ SVC{\isacharunderscore}now{\isacharsemicolon}{\isacharsemicolon}\isanewline
\ \ \ \ \ \ SVC\isactrlsub aEnable{\isacharsemicolon}{\isacharsemicolon}\isanewline
\ \ \ \ \ \ WHILE\ {\isasymnot}SVC\isactrlsub aReq\isanewline
\ \ \ \ \ \ DO\isanewline
\ \ \ \ \ \ \ \ SKIP\isanewline
\ \ \ \ \ \ OD{\isacharparenright}{\isacharparenright}\isanewline
\ \ \ \ OD\isanewline
\ \ \ \ FI\isanewline
\ \ COEND{\isacharparenright}{\isacharparenright}{\isachardoublequoteclose}\isanewline
\isadelimtheory
\endisadelimtheory
\isatagtheory
\endisatagtheory
{\isafoldtheory}%
\isadelimtheory
\endisadelimtheory
\end{isabellebody}%

\paragraph{Acknowledgements} NICTA is funded by the Australian Government through the Department of Communications and the Australian Research Council through the ICT Centre-of-Excellence Program.

\end{document}